\documentclass[8pt,preprint2]{aastex}

\shorttitle{Excess of Neutral Iron K-shell Line in GRXE}
\shortauthors{Nobukawa et al.}

\begin{document}
\title{Enhancement of the 6.4 keV line in the inner Galactic ridge: \\Proton-induced fluorescence?}
\author{K. K. Nobukawa\altaffilmark{1}, M. Nobukawa\altaffilmark{2,1}, H. Uchiyama\altaffilmark{3}, T. G. Tsuru\altaffilmark{1}, K. Torii\altaffilmark{4}, T. Tanaka\altaffilmark{1}, \\D. O. Chernyshov\altaffilmark{5}, Y. Fukui\altaffilmark{4}, V. A. Dogiel\altaffilmark{5,6}, and K. Koyama\altaffilmark{1,7}}

\altaffiltext{1}{Department of Physics, Graduate School of Science, Kyoto University, Kitashirakawa Oiwake-cho, Sakyo-ku, Kyoto 606-8502, Japan.}
\altaffiltext{2}{The Hakubi Center for Advanced Research, Kyoto University, Yoshida-Ushinomiya-cho, Sakyo-ku, Kyoto 606- 8302, Japan.}
\altaffiltext{3}{Faculty of Education, Shizuoka University, 836 Ohya, Suruga-ku, Shizuoka 422-8529, Japan.}
\altaffiltext{4}{Department of Physics, Nagoya University, Chikusa-ku, Nagoya, Aichi 464-8601, Japan.}
\altaffiltext{5}{P. N. Lebedev Institute, Leninskii pr, 53, Moscow 119991, Russia.}
\altaffiltext{6}{Moscow Institute of Physics and Technology, 141700 Moscow Region, Dolgoprudnii, Russia.}
\altaffiltext{7}{Department of Earth and Space Science, Graduate School of Science, Osaka University, 1-1 Machikaneyama-cho, Toyonaka, Osaka 560-0043, Japan.}

\begin{abstract}
A common idea for the origin of the Galactic diffuse X-ray emission, particularly that of the iron lines from neutral and highly ionized atoms, 
is a superposition of many cataclysmic variables and coronally active binaries. In this scenario, the flux should symmetrically distribute between 
the east and west on the plane with respect to Sagittarius A$^*$ because the stellar mass distribution determined by infrared observations is nearly symmetric.  
This symmetry is confirmed for the highly ionized iron line as well as the continuum emission. However, a clear excess of the neutral iron line 
in the near east of the Galactic center compared to the near-west side is found.  The flux distribution of the excess emission well correlates 
with molecular column density. The X-ray spectrum of the excess emission is described by a power-law continuum plus a 6.4~keV line 
with the large equivalent width of $\sim1.3$~keV, 
which is hardly explained by the low-energy electron bombardment scenario. 
The longitudinal and latitudinal distribution of the excess emission disfavors the X-ray irradiation, neither by Sagittarius A$^*$ nor by nearby X-ray binaries.
Then the low-energy proton bombardment is the most probable origin although the high energy density $\sim 80$~eV~cm$^{-3}$ in 0.1--1000~MeV 
is required and there is no conventional proton source in the vicinity.
\end{abstract}


\section{Introduction}
K-shell iron lines are one of the most remarkable features in the Galactic diffuse X-ray emission 
(Galactic center X-ray emission: GCXE and the Galactic ridge X-ray emission: GRXE).  
The spatial distribution of the iron lines largely extends out to $|l|\sim60$\degr~along the plane. 
A deep observation of the GRXE near the Galactic center 
($l=0{\degr}.08$, $b=-1{\degr}.42$)  
has resolved almost all the GRXE flux into point sources \citep{Re09}. 
This fact demonstrates that the origin of the iron lines is due to a superposition of many point sources 
such as cataclysmic variables (CVs) and coronally active binaries (ABs) \citep{Re06, Yu12,Wa14}.  
The spectroscopic studies with {\it ASCA}, {\it XMM-Newton}, {\it Chandra}, and {\it Suzaku} revealed that 
the K-shell iron lines are separated into three lines at 6.4 (neutral iron), 6.7 (He-like iron), and 7.0 keV  
\citep[H-like iron; ][]{Ko96,Pr03,Mu04,Uc13}. 
The X-ray spectra of the GRXE and GCXE, however, are not the same, which indicates different origins \citep{Uc13}. 
Even in the point-source scenario, the spectra and the mixing ratio of CVs and ABs should be different between the GRXE and GCXE. 
Furthermore, the flux distribution of the 6.4 keV line in the GCXE is clumpy, while those of the 6.7 and 7.0~keV lines show a 
roughly smooth and east-west symmetric distribution with respect to Sagittarius (Sgr) A$^*$ in the Galactic center \citep{Uc11}.
 
The 6.4 keV clumps in the GCXE have a large equivalent width (EW $\sim1.0$~keV) and are time variable \citep[e.g.,][]{In09, Po10, No11}, 
and hence their origin is believed to be X-ray reflection and fluorescence by external X-rays of past big flares of Sgr A* 
\citep[X-ray reflection nebulae;][]{Ko96}.  
In addition to the clumps, a more uniformly distributed 6.4 keV line emission is found \citep{Uc11},  
and its origin is under debate.
This Letter reports the first evidence for a clear east-west asymmetry of the 6.4 keV line emission at 
$|l|=$1\degr.5--3\degr.5 
on the Galactic disk: the east shows an excess over the west. 
Based on the spatial and spectral analysis, we discuss the origin of this excess.
Errors quoted in this paper are at 90\% confidence levels unless otherwise specified.
The distance to the Galactic center is assumed to be 8~kpc.

\section{Observations and data reduction}

We performed campaign observations of the inner Galactic disk near the Galactic center at $-4{\degr}<l<-2{\degr}$ (west side) 
and $2{\degr}< l < 4{\degr}$ (east side) using the {\it Suzaku} X-ray observatory \citep{Mi07}. 
The observation log is listed in table~1.
The X-ray charge coupled device camera, the XIS \citep{Ko07a}, onboard {\it Suzaku} covers the energy range of 0.2--12~keV, 
and has the field of view of $17'.8\times17'.8$. We excluded events during the South Atlantic Anomaly passages, 
at elevation angles below $3{\degr}$ from the night Earth rim, and at elevation angles below $10{\degr}$ from the sunlit Earth rim. 
We further excluded the data of $\sim3$~ks during background flares detected on 17 March 2013. 
We reprocessed the data by using {\tt xispi} in the analysis software package, HEAsoft 6.15.1, and the {\it Suzaku} 
calibration database ({\tt CALDB}) released in May 2014. After the screening, we obtained the total exposure times of 
597~ks and 1187~ks for the west and east data sets, respectively.

A bright X-ray binary, GX\,3$+$1, is located at $(l, b)=(2\degr.294, 0\degr.794)$, 
which is $30'$--$90'$ apart from each pointing in the east observations. 
Since some regions of the XIS fields of view are contaminated by the stray light from GX\,3$+$1,  we used data in the other clean regions.
We excluded circular regions with $2''$ radii at the positions of point-like sources with fluxes 
higher than $1\times10^{-13}$~erg~s$^{-1}$~cm$^{-2}$ in 2--10~keV. We, furthermore, detected  two compact diffuse sources 
in the east and west regions.  While the west source is identified as G357.7$-$0.1 
(Tornado nebula: \citealt{Sa11}), no counterpart is found for the east source, which is located at $(l, b)=(3{\degr}.337, -0{\degr}.055)$.  
These two sources are also excluded from the data.

\begin{table*}
\begin{center}
\caption{Observation log.}
\begin{tabular}{crrcc}
\tableline\tableline
Obs. ID 	& \multicolumn{2}{c}{Pointing direction} 	& Obs. start 			& Exposure time\tablenotemark{a}	\\
			& $l~({\degr})$		& $b~({\degr})$			& (UT)					& (ks)		\\
\tableline
\multicolumn{5}{l}{Western side} \\
\tableline
501052010 	& -1.5004 			& -0.0034 				& 2006-10-10 06:45:09 	& 21.0		\\
501053010	& -1.8335 			& -0.0030 				& 2006-10-10 21:18:59 	& 23.4		\\
503014010 	& -2.0995 			& -0.0522 				& 2008-09-18 04:46:49 	& 59.7		\\
504036010 	& -2.2943 			& -0.1163 				& 2009-08-29 12:05:20 	& 136.5		\\
503015010 	& -2.3502 			& -0.0527 				& 2008-09-19 07:33:05 	& 61.4		\\
503016010 	& -2.6012 			& -0.0525 				& 2008-09-22 06:47:49 	& 57.4		\\
503017010 	& -2.8503 			& -0.0525 				& 2008-09-23 08:08:10 	& 56.5		\\
503018010 	& -3.1015 			& -0.0517 				& 2008-09-24 09:27:54 	& 31.9		\\
503018020 	& -3.1004 			& -0.0514 				& 2008-10-03 18:05:13 	& 13.3		\\
503018030 	& -3.1006 			& -0.0470 				& 2009-02-19 07:32:01 	& 12.9		\\
503019010 	& -3.3496 			& -0.0477 				& 2009-02-19 16:37:49 	& 56.8		\\
503020010 	& -3.6001 			& -0.0473 				& 2009-02-21 01:15:55 	& 66.2		\\
\tableline
\multicolumn{5}{l}{Eastern side} \\
\tableline
501060010 	& 1.5016 			& 0.0027 				& 2007-03-17 05:07:04 	& 68.1		\\
508075010 	& 1.7513 			& -0.0431				& 2014-03-10 01:33:32 	& 109.3		\\
502009010 	& 1.8336 			& -0.0035 				& 2007-10-12 21:52:24 	& 22.9		\\
507069010 	& 2.0003 			& -0.0439 				& 2013-03-15 09:48:19 	& 110.3		\\
507070010 	& 2.2511 			& -0.0437 				& 2013-03-17 18:39:56 	& 111.8		\\
507071010 	& 2.5011 			& -0.0438 				& 2013-03-20 02:41:04 	& 112.3		\\
507072010 	& 2.7509 			& -0.0439 				& 2013-03-22 07:20:36 	& 110.7		\\
507073010 	& 3.0011 			& -0.0434 				& 2013-03-24 08:46:03 	& 108.9		\\
507074010 	& 3.1514 			& 0.1563 				& 2013-04-03 21:33:46 	& 104.1		\\
507075010 	& 3.2512 			& 0.4070 				& 2013-03-11 09:06:19 	& 109.6		\\
508076010 	& 3.2514 			& -0.0427 				& 2014-02-28 12:46:16 	& 109.8		\\
508077010 	& 3.5043 			& -0.0380 				& 2014-03-02 17:00:51	& 109.4		\\
\tableline
\end{tabular}
\tablenotetext{a}{Effective exposure time after the screening (see text).}
\end{center}
\end{table*}

\section{Results and discussion}
\subsection{Galactic longitude profile} \label{profile}

We make spectra from each pointing and subtract the non X-ray background (NXB) 
estimated with 
 {\tt xisnxbgen} \citep{Ta08}.
The NXB-subtracted spectra in the 4--10~keV band are fitted with a phenomenological model which  consists of four Gaussian lines and an absorbed bremsstrahlung as well as the cosmic X-ray background (CXB) model \citep{Ku02}. 
The line energies of the Gaussians are fixed to 6.4 (neutral iron K$\alpha$), 6.7 (He-like iron K$\alpha$), 7.0 (H-like iron Ly$\alpha$), 
and 7.06~keV \citep[neutral iron K$\beta$; ][]{Sm01,Ka93}.
The intensity of the neutral iron K$\beta$ line is fixed to 0.125 times that of the neutral iron K$\alpha$ line 
 \citep{Ka93}, 
while those of the other lines are free parameters. The interstellar absorption column densities for the east 
and the west are fixed to $5\times10^{22}$~cm$^{-2}$ \citep{Uc13}. The normalization and temperature of the bremsstrahlung component  
are allowed to vary. We then  obtain the intensities of the 6.4~keV and 6.7~keV lines and the continuum in the 4--10~keV band (where the 6--7~keV band was excluded).

The flux of the GRXE depends on  the Galactic latitude \citep{Ka97,Uc13}. The geometric centers of the fields of view 
are not exactly on the Galactic plane ($b = -0{\degr}.046$), 
and hence we correct the best-fit intensities to those on the Galactic plane. 
We use the scale-height of $0{\degr}.7$ for the GRXE \citep{Ka97}. 
Then the relation between the best-fit intensity $I_{\rm obs}$ at $b$ 
and the corrected intensity $I_{\rm cor}$ is:

\begin{equation}
I_{\rm obs}={\rm exp}\,\Biggl(-\frac{|\,b+0^{\circ}\!.046\,|}{0^{\circ}\!.7}\Biggr) \times I_{\rm cor}\,.
\end{equation}

The longitudinal profile of the intensity of the 6.4~keV and 6.7~keV lines and the continuum are shown in figure~1. 
Previous studies revealed that a major fraction of the GRXE
is due to unresolved point sources, mostly CVs and ABs \citep{Re06,Re09,Yu12,Wa14}. 
The stellar distribution has been constructed from infrared observations, and has an east-west symmetry \citep{Ni13,La02,Re06}.  
The profile is given by the dashed lines in figure~1. All the X-ray flux distributions, except that of the 6.4 keV line in the east, 
are consistent with the stellar distribution curve. A remarkable structure is an excess of the 6.4 keV line flux in the east compared to the west, and hence the excess should have a different origin from the symmetrical components due to point sources.

We plot in figure~1 the $^{12}$CO line intensity taken by {\it NANTEN} 
integrated over the velocity range from $-300$~km~s$^{-1}$ to $+300$~km~s$^{-1}$ \citep{To10}. 
Here, the local components uniformly distributed around $-20$~km~s$^{-1}$ to $+30$~km~s$^{-1}$ and the near-side 3~kpc arm having a velocity gradient 
from $-80$~km~s$^{-1}$ at $l=-5$\degr~to $-30$~km~s$^{-1}$ at $l=5$\degr~\citep{Da08} are excluded.
The $^{12}$CO profile is similar to the excess distribution of the 6.4~keV line in the east,  which suggests that its origin is due to molecular gas. 
The excess at $l = 3{\degr}$ coincides with 
the intersection point of the Galactic plane and the giant molecular cloud, Bania's Clump 2, which extends towards the north \citep{Ba77}. 

\begin{figure*}[tb]
\epsscale{1.4}
\plotone{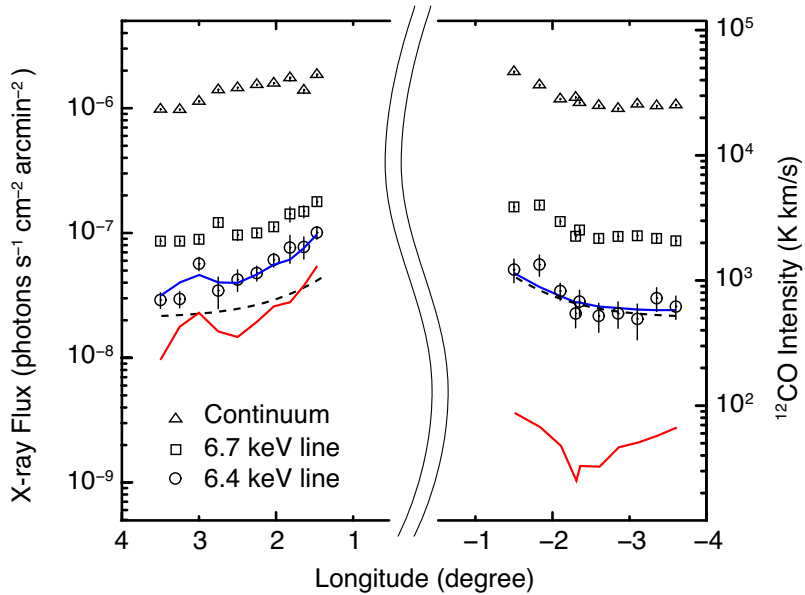}
\caption{Intensity profiles of X-rays and $^{12}$CO molecular clouds on the Galactic plane ($b = -0{\degr}.046$). 
The triangles, squares, and circles show the fluxes of the continuum in the 4--10~keV band (where the 6--7~keV band was excluded), 
the 6.7~keV line, and the 6.4~keV line, respectively. Errors are quoted at 68\% confidence levels. 
The dashed lines show a model of stellar distribution, which is symmetric with respect to Sgr~A$^{*}$ \citep{Ni13,La02,Re06}. 
The red lines are the  $^{12}$CO intensity profile (the unit is the right-side axis). 
The blue lines are  the sum of the symmetric distribution model and the $^{12}$CO intensity multiplied by $\alpha$, where $\alpha$ is 
a factor to convert the $^{12}$CO intensity to the 6.4~keV flux (see text).}
\end{figure*}

\subsection{Spectrum of the excess of the 6.4~keV line in the east} \label{spectrum}

\begin{figure*}[tb]
\begin{center}
\includegraphics[width=15cm]{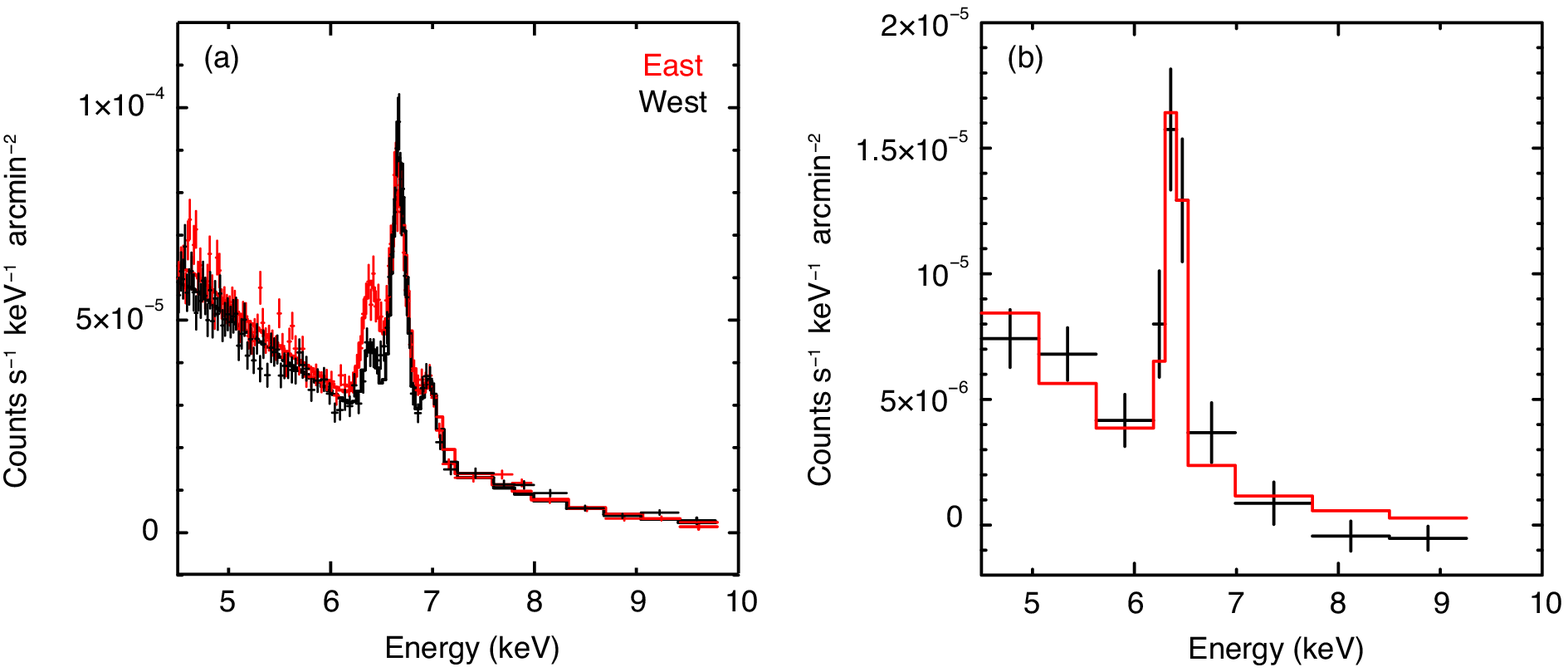}
\caption{(a) X-ray spectra extracted from the east (red) and west (black) sides.  The red and black lines show 
the best-fit model consisting of an absorbed power law plus four Gaussian lines at 6.4, 6.7, 7.0, and 7.06 keV. 
(b) X-ray spectrum of the excess emission in the east. The red line shows 
the best-fit model consisting of a power law plus a Gaussian line at 6.4~keV.
Errors are quoted at 68\% confidence levels for the both panels.}
\end{center}
\end{figure*}

\begin{table*}[tb]
\begin{center}
\caption{Best-fit parameters of the X-ray spectra from the east and west sides.\tablenotemark{a}}
\begin{tabular}{cccccc}
\tableline\tableline
		 	& \multicolumn{3}{c}{Emission lines} 															& \multicolumn{2}{c}{Continuum}	\\
\tableline
			& 6.4 keV						& 6.7 keV						& 7.0 keV						& photon index		& normalization	\\
Unit		& ($10^{-8}$)\tablenotemark{b}	& ($10^{-8}$)\tablenotemark{b}	& ($10^{-8}$)\tablenotemark{b}	&					& ($10^{-8}$)\tablenotemark{c}		\\
\tableline
EAST	 	& 5.1$\pm$0.4		 			& 10.3$\pm$0.5	 				& 2.5$\pm$0.4 					& 2.3$\pm$0.1		& 22.9$\pm$0.1		\\
WEST		& 2.5$\pm$0.4		 			& 10.3$\pm$0.5	 				& 2.8$\pm$0.4 					& 2.2$\pm$0.1		& 21.7$\pm$0.1		\\
excess 	 	& 2.5$\pm$0.6 					& --	 						& --						 	& 3$\pm$1			& 1.9$\pm$0.3		\\
\tableline
\end{tabular}
\tablenotetext{a}{The error ranges in this table are calculated at 90\% confidence levels.}
\tablenotetext{b}{Absorption-corrected line flux in the units of photons~s$^{-1}$~cm$^{-2}$~arcmin$^{-2}$.}
\tablenotetext{c}{Normalization at 6.4 keV in the units of photons~s$^{-1}$~keV$^{-1}$~cm$^{-2}$~arcmin$^{-2}$.}
\end{center}
\end{table*}

Figure~2a shows integrated X-ray spectra from the east and west sides. The CXB is subtracted from both the spectra according to \cite{Ku02}. 
Each spectrum is fitted with an absorbed power law plus four Gaussian lines at 6.4, 6.7, 7.0, and 7.06~keV.  
The interstellar absorption column density is fixed to $5\times10^{22}$~cm$^{-2}$. 
The intensity of 7.06~keV (the neutral iron K$\beta$) line is fixed to 0.125 times that of 6.4~keV (the neutral iron K$\alpha$) line.
The best-fit parameters are summarized in table~2. 

We subtract the west spectrum from the east one to make the X-ray spectrum of 
the east excess as is shown in figure~2b. 
The excess spectrum is fitted with a power law plus a Gaussian line at 6.4~keV. 
The best-fit continuum and the line fluxes are $(1.9\pm0.3)\times10^{-8} 
(E/6.4~{\rm keV})^{-3\pm1}$~photons s$^{-1}$ keV$^{-1}$ cm$^{-2}$~arcmin$^{-2}$ 
and $(2.5\pm0.6)\times10^{-8}$~photons s$^{-1}$ cm$^{-2}$ arcmin$^{-2}$, respectively, as shown in table~2. 
Since the 6.7~keV line and its relevant continuum fluxes have a possible asymmetry between the east and west by $\pm7$\%, 
we take this error into account as the uncertainty of the symmetry.  Then the EW is estimated to be  
$1.3\pm0.4^{+4.2}_{-0.2}$~keV, where the second and third terms are the statistical and systematical errors, respectively.

In the case of cosmic-ray bombardment, the 6.4~keV line is produced via inner-shell ionization by protons in the MeV band \citep{Do11} or 
by electrons in the keV band \citep{Va00,Yu02}, while the continuum is due to inverse bremsstrahlung (for protons) 
or bremsstrahlung (for electrons). 
The photon index of the observed X-ray spectrum ($\Gamma = 3\pm1$) translates into the particle index $2.5\pm1.0$.
As for the X-ray irradiation, the 6.4 keV line is produced via photoionization by X-rays with energy higher than 
7.1~keV (K-edge) while the continuum is due to Thomson scattering. 
The photon index of the observed X-ray spectrum is the same as that of the irradiating X-rays. 
We show the EW of the 6.4~keV line as a function of particle or photon index in figure~3. 
In both the scenarios, the EW depends on the iron abundance \citep{Ts07,Do11}.  
The metal abundances were determined in the Galactic center region by X-ray observations 
of the high temperature plasma to be  $\sim1.9$~solar for sulfur, argon, calcium 
but $\sim$1.2~solar for iron \citep{No10,Uc13}. \cite{Mu04} reported a similar result. 
The iron abundance of 1--1.5 solar was also measured in the X-ray reflection nebulae 
in Sgr B with the absorption of the Fe-K edge \citep{No11}. 
The same result was obtained by mid-infrared observations; 
\cite{Gi02} obtained 2~solar for heavy elements of neon, sulfur, and argon (not of iron), and 
\cite{Cu07} and \cite{Ma15} measured the iron abundance to be 1--1.5~solar. Therefore, we adopt the iron abundance of 1--1.5~solar in figure~3.
The EW in the X-ray scenario also depends on the angle $\theta$ between the line of sight and the incident photon direction \citep{Ts07}. 
Figure~3 shows the result of the reflection angle $\theta=90^{\circ}$, which gives the maximum EW.

The black horizontal line in figure~3 indicates the best-fit EW, while the horizontal hatched region shows 
the range of the statistical and systematic errors. 
The allowed parameter region completely excludes the electron origin. 
Then, we discuss the others, the X-ray and proton scenarios in the following subsections.

\begin{figure}[tb]
\begin{center}
\includegraphics[width=7.7cm]{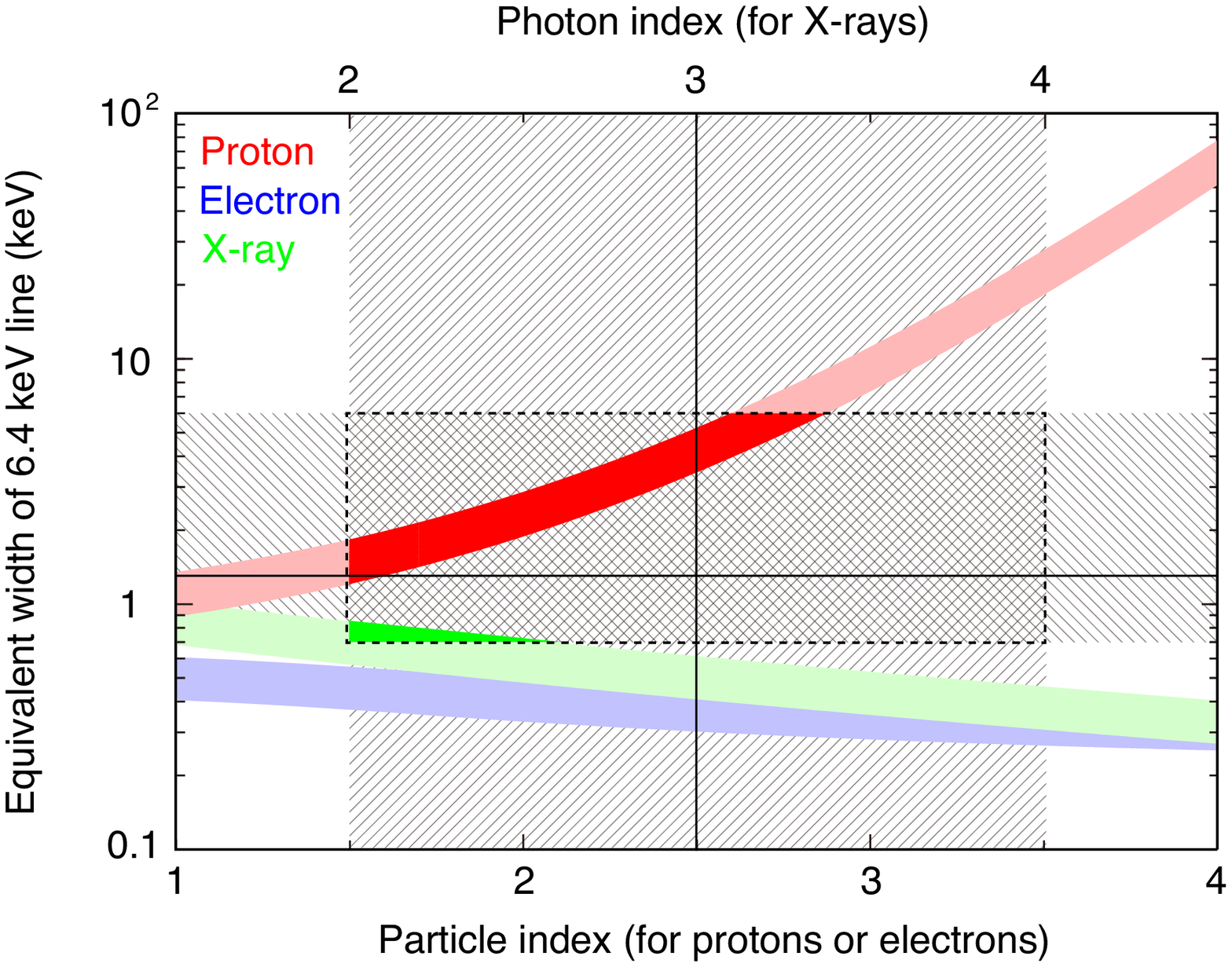}
\caption{EW of the 6.4 keV~line as a function of particle index \citep{Do11} or photon index \citep{Ts07} with the iron abundance of 1--1.5 solar (see text). The thin red, blue and green belts are  the calculated values for protons, electrons, and X-rays, respectively. 
The best-fit value and error region of the photon or particle index are shown by the vertical line and hatched region, respectively. The thick red and green regions are 
the acceptable ranges for the proton and X-ray scenarios, respectively.}
\end{center}
\end{figure}

\subsection{X-ray irradiation scenario}
In  the X-ray irradiation scenario, the 6.4 keV line flux depends on the irradiating source flux, 
the hydrogen column density ($N_{\rm H}$), and the distance ($D$) between the irradiating source and the target.
The hydrogen column density $N_{\rm H}=$(2--6)$\times10^{22}$~cm$^{-2}$ 
is obtained by multiplying the $^{12}$CO line intensity in figure~1 by the conversion factor (called $X$-factor) of 
$0.7\times10^{20}$~cm$^{-2}$~(K$\,\cdot\,$km~s$^{-1}$)$^{-1}$ \citep{To10}.

A possible irradiating source is the supermassive black hole Sgr A$^*$, as for the X-ray reflection nebula model invoked for the GC region 
\citep{Su93,Ko96,Po10}.
In this case, the source luminosity should be $L_{\rm X} \sim 10^{40}~(D/450~{\rm pc})^{2}$~erg~s$^{-1}$ about $1500\times (D/450~{\rm pc})$ years ago, where $D$ is the distance of the 6.4 keV line emitting region from Sgr A$^*$.   
For the X-ray reflection nebulae in Sgr B, the luminosity of Sgr A$^*$ was $\sim 10^{39}$~erg~s$^{-1}$ about 300 years ago \citep{Mu00, Re04}. 
Brighter flares of $\sim 10^{41-43}$~erg~s$^{-1}$ in the far past (10$^5$--10$^7$ years ago) have been proposed to explain 
the Fermi bubbles 
and the recombining plasma in the south region of the Galactic center \citep{Su10,Na13}.  
Thus, many big flares in the interval period between these two epochs 
($\sim$700--1500 years ago) 
may be conceivable.  
However, this scenario has two difficulties.

The first difficulty comes from the longitudinal distribution of the excess flux of the 6.4 keV line. 
The observed excess flux is proportional to the $^{12}$CO intensity, 
which means that 
the X-ray intensity from Sgr A$^*$ is almost constant, and does not decrease as the square of the  Galactic longitude, or the distance from Sgr A$^*$;  
the past Sgr A$^*$ flares should have smoothly increasing flux with the square of the look-back time. 
A monotonous smooth decrease of flux should continue during 700--1500 years ago, 
which is artificial, although not completely rejected.

The second difficulty comes from the flux ratio of the 6.4 keV line to the $^{12}$CO intensity (here, $\alpha$) in the molecular cloud Clump~2.  
Clump~2 extends toward the north from the Galactic plane and has an elliptical shape with major and minor axes of 
$\sim1{\degr}.1$ and $\sim0{\degr}.5$, respectively.
We estimate the excess 6.4 keV line flux from the on-plane  
($b\sim-0{\degr}.04$; Obs. ID=507073010) 
and off-plane 
($b\sim0{\degr}.15$--0{\degr}.40; Obs.ID=507074010 and 507075010) 
parts of Clump~2 
by subtracting the symmetrical component from the observed flux, 
and obtained $\alpha$ of $(5.9\pm0.7)\times10^{-11}$ and $(2.1\pm0.6)\times10^{-11}$ for the on- and off-plane parts, respectively (unit is photons~s$^{-1}$~cm$^{-2}$~arcmin$^{-2}$ (K$\,\cdot\,$km~s$^{-1}$)$^{-1}$). 
Since the flux from Sgr A$^*$ would be equal in the small separation angle between  on- and off-plane parts, 
$\alpha$ should be also equal, which is in conflict with the observed values.
From these two difficulties, we regard that the X-ray irradiation by Sgr A$^*$ is unlikely. 

The other possibility is that irradiation of many X-ray binaries in the east is responsible for the 6.4~keV line. 
This idea is essentially the same as \cite{Mo14}. 
Assuming that a mean spectrum of the relevant X-ray binaries is a power law with the photon index of $\Gamma=3$, 
and is surrounded by cold material with $N_{\rm H}=2$--$6\times10^{22}$~cm$^{-2}$, 
a fraction of 0.03--0.1\% of the total 2--10 keV flux is converted to 6.4~keV X-rays. 
We estimate the excess 6.4~keV luminosity in total to be $3\times10^{33}$~erg~s$^{-1}$ in the relevant area of $2\degr.0 \times 0\degr.2$.  
Then, a total luminosity of $3\times10^{36}$--$10^{37}$~erg~s$^{-1}$ is required to the sources inside the area.
This flux exceeds the total luminosity of the GRXE in this region ($\sim 3\times10^{35}$~erg~s$^{-1}$ ) by one or two order of magnitude.  No source brighter than $10^{35}$~erg~s$^{-1}$ is found in this area. 
Therefore, more than one or two order of magnitude brighter flares within the relevant regions and duration of the past $\sim 1000$ years are required. 
In addition, the reflection angle $\theta$ should be distributed randomly. 
Then the EW integrated over the reflection angle $\theta$ is $\sim1.3$~times lower than the green belt in figure~3 (reflection angle $\theta=90^{\circ}$), and hence is out of the allowed range.  
Thus this scenario is also unlikely. 

\subsection{Energy density of low-energy protons} \label{density}
In the proton model, 
the excess flux of the 6.4 keV line $I_{\rm 6.4\,keV}$ is given by 
\begin{equation}
{\tiny
I_{\rm 6.4\,keV} = \frac{1}{4 \pi}\ N_{\rm H} \int \sigma_{\rm 6.4\, keV}\ \upsilon\ A \left( \frac{E_{\rm p}}{1~{\rm MeV}}\right)^{-2.5}\ dE_{\rm p},
}
\end{equation}
where $\sigma_{\rm 6.4~keV}$, $\upsilon$, $A$, $E_{\rm p}$
and $N_{\rm H}$ are the cross section to produce the 6.4 keV line by protons, the velocity,
the number density at 1~MeV and the energy of protons, 
and the line-of-sight hydrogen column density, respectively. 
The spectral index of $-2.5$ is adopted from the best-fit result (see figure~3).
Using the $X$-factor of $0.7\times10^{20}$~cm$^{-2}$~(K$\,\cdot\,$km~s$^{-1}$)$^{-1}$ \citep{To10}, 
the 6.4 keV line flux is expressed as the $^{12}$CO line intensity 
multiplied by the conversion factor $\alpha$, which is measured to be 
$4.2\times10^{-11}$~photons~s$^{-1}$~cm$^{-2}$~arcmin$^{-2}$~(K$\,\cdot\,$km~s$^{-1}$)$^{-1}$ in figure~1. 
The cross section $\sigma_{\rm 6.4\,keV}$ has a peak at 10~MeV and rapidly decreases below 1~MeV and above 50~MeV \citep{Pa89}.
We set the integration range to be 1--50~MeV, and then
obtained the normalization $A=1.4 \times 10^{-5}$~protons~cm$^{-3}$ and the energy density of 20~eV~cm$^{-3}$.  
When the integration range is expanded to be 0.1--1000~MeV, the energy density becomes 80~eV~cm$^{-3}$.
This is about one or two orders of magnitude higher 
than the canonical value $\sim1$~eV~cm$^{-3}$ 
that is determined by observing high-energy cosmic rays \citep{Ne12}. 

\cite{Ta12} calculated an energy conversion rate from protons to the 6.4 keV line to be $10^{-6}$ or less. 
Since the total luminosity of the 6.4 keV line emission is $3\times 10^{33}$~erg~s$^{-1}$, the proton power of $>3\times10^{39}$~erg~s$^{-1}$ is estimated. 
This is not far from the energy $\sim2\times10^{39}$~erg~s$^{-1}$ that protons input to the Galactic center \citep{Do13}. 

The diffusion length of low-energy ($\sim$MeV) protons is only a few tens of parsecs \citep{Do11}, and therefore the MeV protons 
should be produced {\it in situ}, possibly by a supernova remnant 
or 
a pulsar wind nebula. 
However, no candidate source is found in the vicinity.
\cite{Am11} indicated that cosmic-ray particles are possibly generated with stochastic acceleration 
by Alfv\'enic turbulence in the central molecular zone with large velocity dispersion of $\sim100$~km~s$^{-1}$. 
Since Clump~2 exhibits large velocity dispersion of $\sim 100$~km~s$^{-1}$ \citep{Ba77,To10}, 
another possibility to produce the MeV protons is the stochastic acceleration.
Our results demonstrate that 
the 6.4 keV line can be a unique probe to investigate low-energy cosmic-ray protons. 
\acknowledgments

The authors thank all the members of the Suzaku team and the NANTEN consortium. K.K.N. is supported by Research Fellowships of JSPS for Young Scientists. This study was also supported by JSPS and MEXT KAKENHI Grant Numbers 24740123 (M.N.), 25887028 (H.U.), 23340047, 25109004, 15H02090 (T.G.T.), 23740149 (K.T.), and 24540229 (K.K.). V.A.D. and D.O.C acknowledge a partial support from the RFFI grants 15-52-52004 and 15-02-02358. D.O.C. is partially supported by the Dynasty foundation.



\begin{thebibliography}{}
\bibitem[Amano et al.(2011)]{Am11}
Amano, T., Torii, K., Hayakawa, T., \& Fukui, Y. 2011, PASJ, 63, L63
\bibitem[Bania(1977)]{Ba77}
Bania, T. M. 1977, \apj, 216, 381
\bibitem[Cunha et al.(2007)]{Cu07}
Cunha, K., Sellgren, K., Smith, V. V., et al. 2007, \apj, 669, 1011
\bibitem[Dame and Thaddeus(2008)]{Da08}
Dame, T. M. \& Thaddeus, P. 2008, ApJL, 683, L143
\bibitem[Dogiel et al.(2011)]{Do11}
Dogiel, V., Chernyshov, D., Koyama, K., Nobukawa, M., \& Cheng, K. 2011, PASJ, 63, 535
\bibitem[Dogiel et al.(2013)]{Do13}
Dogiel, V. A., Chernyshov, D. O., Tatischeff, V., Cheng, K.-S., \& Terrier, R. 2013, ApJL, 771, 43
\bibitem[Giveon et al.(2002)]{Gi02}
Giveon, U., Sternberg, A., Lutz, D., Feuchtgruber, H., \& Pauldrach, A. W. A. 2002, ApJ, 566, 880
\bibitem[Inui et al.(2009)]{In09}
Inui, T., Koyama, K., Matsumoto, H., \& Tsuru, T. G. 2009, PASJ, 61, S241
\bibitem[Kaastra and Mewe(1993)]{Ka93}
Kaastra, J. S. \& Mewe, R. 1993, A\&A, 97, 443
\bibitem[Kaneda et al.(1997)]{Ka97}
Kaneda, H., Makishima, K., Yamauchi, S., et al. 1997, \apj, 491, 638
\bibitem[Koyama et al.(1996)]{Ko96}
Koyama, K., Maeda, Y., Sonobe, T., et al. 1996, PASJ, 48, 249
\bibitem[Koyama et al.(2007)]{Ko07a}
Koyama, K., Tsunemi, H., Dotani, T., et al. 2007, PASJ, 59, S23
\bibitem[Kushino et al.(2002)]{Ku02} 
Kushino, A., Ishisaki, Y., Morita, U., et al. 2002, PASJ, 54, 327
\bibitem[Launhardt et al.(2002)]{La02}
Launhardt, R., Zylka, R., \& Mezger, P. G. 2002, A\&A, 384, 112
\bibitem[Martin et al.(2015)]{Ma15} 
Martin, R. P., Andrievsky, S. M., Kovtyukh, V. V., et al. 2015, MNRAS, 449, 4071
\bibitem[Mitsuda et al.(2007)]{Mi07} 
Mitsuda, K., Bautz, M., Inoue, H., et al. 2007, PASJ, 59, S1
\bibitem[Molaro et al.(2014)]{Mo14}
Molaro, M., Khatri, R., \& Sunyaev, R. A. 2014, A\&A, 561, 107
\bibitem[Muno et al.(2004)]{Mu04} 
Muno, M. P., Baganoff, F. K., Bautz, M. W., et al. 2004, \apj, 613, 326
\bibitem[Murakami et al.(2000)]{Mu00}
Murakami, H., Koyama, K., Sakano, M., Tsujmoto, M., \& Maeda, Y. 2000, \apj, 534, 283
\bibitem[Nakashima et al.(2013)]{Na13}
Nakashima, S., Nobukawa, M., Uchida, H. et al. 2013, ApJ, 773, article id. 20
\bibitem[Neronov et al.(2012)]{Ne12}
Neronov, A., Semikoz, D. V., \& Taylor, A. M. 2012, Phys. Rev. Lett. 108, 051105
\bibitem[Nishiyama et al.(2013)]{Ni13}
Nishiyama, S., Yasui, K., Nagata, T., et al. 2013, ApJL, 769, L28
\bibitem[Nobukawa et al.(2010)]{No10}
Nobukawa, M., Koyama, K., Tsuru, T. G., Ryu, S. G., \& Tatischef, V. 2010, PASJ, 62, 423
\bibitem[Nobukawa et al.(2011)]{No11}
Nobukawa, M., Ryu, S. G., Tsuru, T. G., \& Koyama, K. 2011, ApJL, 739, L52
\bibitem[Paul and Sacher(1989)]{Pa89}
Paul, H., \& Sacher, J. 1989, Atomic Data and Nuclear Data Tables, 42, 105
\bibitem[Predehl et al.(2003)]{Pr03}
Predehl, P., Constantini, E., Hasinger, G., \& Tanaka, Y. 2003, Astron. Nachr, 324, 73
\bibitem[Ponti et al.(2010)]{Po10}
Ponti, G., Terrier, R., Goldwurm, A., Belanger, G., \& Trap, G. 2010, \apj, 714, 732
\bibitem[Revnivtsev et al.(2004)]{Re04}
Revnivtsev, M., Churazov, E., Sazonov, M., et al. 2004, A\&A, 425, L49
\bibitem[Revnivtsev et al.(2006)]{Re06}
Revnivtsev, M., Sazonov, S., Gilfanov, M., Churazov, E., \& Sunyaev, R. 2006, A\&A, 452, 169
\bibitem[Revnivtsev et al.(2009)]{Re09}
Revnivtsev, M., Sazonov, S., Churazov, E., et al. 2009, Natur, 458, 1142
\bibitem[Sawada et al.(2011)]{Sa11}
Sawada, M., Tsuru, T. G., Koyama, K., \& Oka, T. 2011, PASJ, 63, S849
\bibitem[Smith et al.(2001)]{Sm01}
Smith, R. K., Brickhouse, N. S., Liedahl, D. A., \& Raymond, L. C. 2001, \apj, 556, L91
\bibitem[Su et al.(2010)]{Su10}
Su, M., Slatyer, T., \& Finkbeiner, D. 2010, ApJ, 724, 1044
\bibitem[Sunyaev et al.(1993)]{Su93}
Sunyaev, R., Markevitch, M., \& Pavlinsky, M. 1993, \apj, 407, 606
\bibitem[Tawa et al.(2008)]{Ta08}
Tawa, N., Hayashida, K., Nagai, M., et al. 2008, PASJ, 60, S11
\bibitem[Tatischeff et al.(2012)]{Ta12}
Tatischeff, V., Decourchelle, A., \& Maurin, G. 2012, A\&A, 546, A88
\bibitem[Torii et al.(2010)]{To10}
Torii, K., Kudo, N., Fujishita, M., et al. 2010, PASJ, 62, 1307
\bibitem[Tsujimoto et al.(2007)]{Ts07}
Tsujimoto, M., Hyodo, Y., \& Koyama, K. 2007, PASJ, 59, S229
\bibitem[Uchiyama et al.(2011)]{Uc11}
Uchiyama, H., Nobukawa, M., Tsuru, T. G., Koyama, K., \& Matsumoto, H. 2011, PASJ, 63, S903
\bibitem[Uchiyama et al.(2013)]{Uc13}
Uchiyama, H., Nobukawa, M., Tsuru, T. G., \& Koyama, K. 2013, PASJ, 65, 19
\bibitem[Valinia et al.(2000)]{Va00}
Valinia, A., Tatischeff, V., Arnaud, K., Ebisawa, K., \& Ramaty, R. 2000, \apj, 543, 733
\bibitem[Warwick et al.(2014)]{Wa14}
Warwick, R. S., Byckling, K., \& P\'erez-Ram\'irez, D. 2014, MNRAS,  438, 2967
\bibitem[Yuasa et al.(2012)]{Yu12}
Yuasa, T., Makishima, K., \& Nakazawa, K. 2012, \apj, 753, 129
\bibitem[Yusef-Zadeh et al.(2002)]{Yu02}
Yusef-Zadeh, F., Law, C., \& Wardle, M. 2002, ApJL, 568, L121

\end{thebibliography}
\end{document}